\documentstyle[psfig,aps]{revtex}

\tightenlines

\begin{document}
\draft

\title{Erratum: Bremsstrahlung from an Equilibrating Quark-Gluon Plasma\\
{[Phys. Rev. C62, 014902 (2000)]}}

\author{Munshi G. Mustafa and Markus H. Thoma}

\date{\today}

\maketitle


Due to a numerical error the rates for bremsstrahlung (4) and annihilation 
with scattering (5) in \cite{Mustafa}, taken from \cite{Aurenche},
were overestimated by a factor of 4 \cite{Steffen}. The quantities
$J_{T,L}$ appearing in (4) and (5) read correctly $J_T=1.11$, $J_L=-1.07$
for two quark flavors $N_f=2$ and $J_T=1.20$, $J_L=-1.13$ for $N_f=3$,
respectively. The photon spectra at RHIC and LHC in equilibrium and chemical
non-equilibrium calculated using the corrected rates are shown in Fig.1-4,
which replace the corresponding figures in \cite{Mustafa}.

The new conclusions following from these figures are the following.
The total photon yield from the quark-gluon plasma at RHIC and LHC is reduced 
by a factor 2-3 compared to \cite{Mustafa}. At RHIC the 
non-equilibrium spectrum is smaller by a factor of 5 at $p_T=1$ GeV, but 
larger by a factor of 2 at 5 GeV, while it is suppressed at LHC
by a factor of about 3 at all momenta compared to the equilibrated case,
if the same initial energy density is assumed. Therefore we conclude
as in \cite{Mustafa} that the equilibrium spectrum is similar to the 
equlibrating. The annihilation-with-scattering contribution to the spectrum 
dominates in equilibrium for all momenta, whereas in the equilibrating case
the one-loop (Compton scattering, anihilation) contribution is 
the largest for all momenta and initial conditions (SSPC, HIJING-I, HIJING-II)
in contrast to the results found in \cite{Mustafa}.
The thermal emission in chemical non-equilibrium exceeds now the prompt 
photon yield for $p_T<3.5$ GeV at RHIC as well as LHC.

\newpage

\begin{figure}
\vspace*{-0.8cm}
\centerline{\psfig{figure=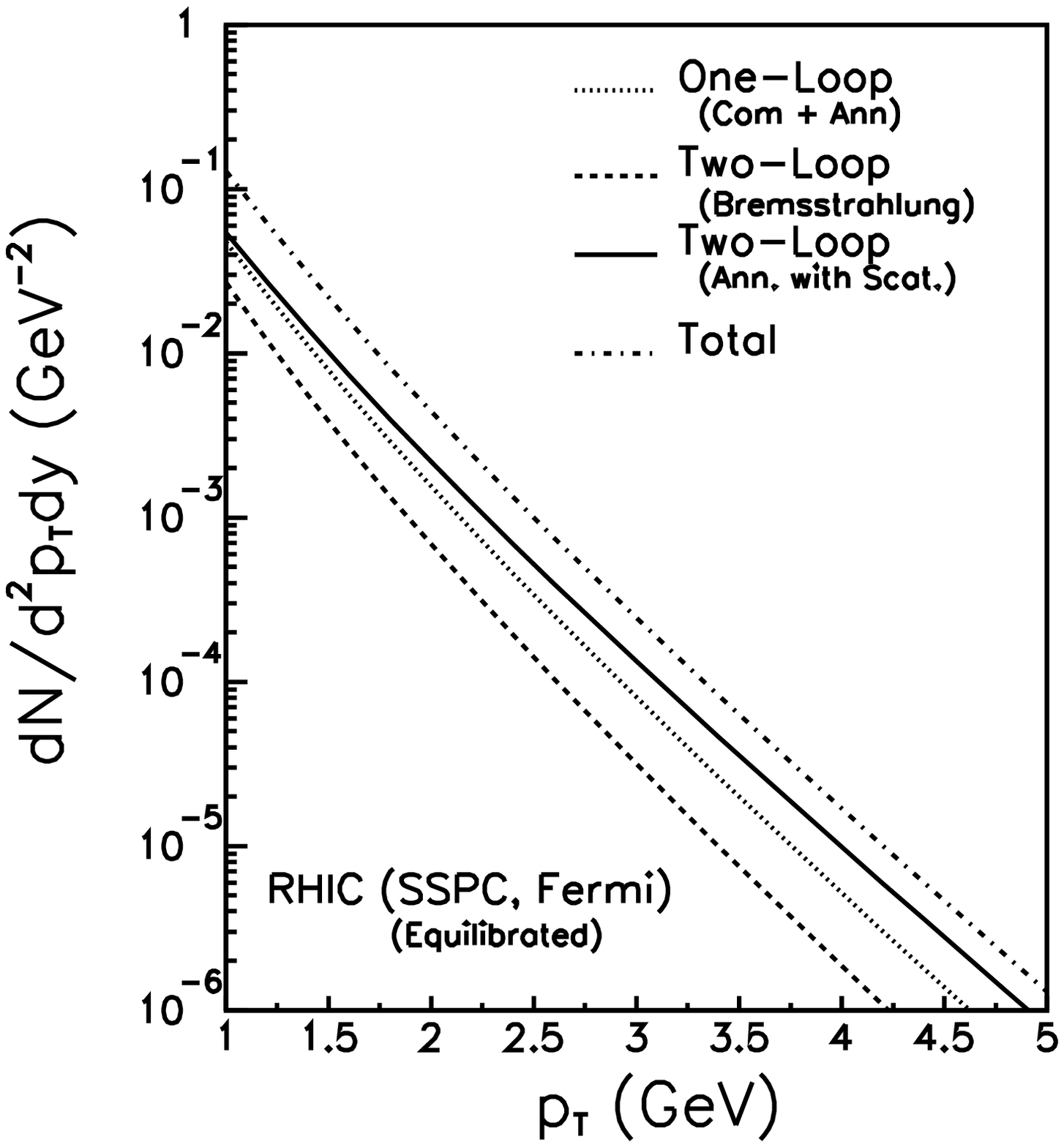,width=14cm,height=11cm}}
\vspace*{-1.9cm}
\centerline{\psfig{figure=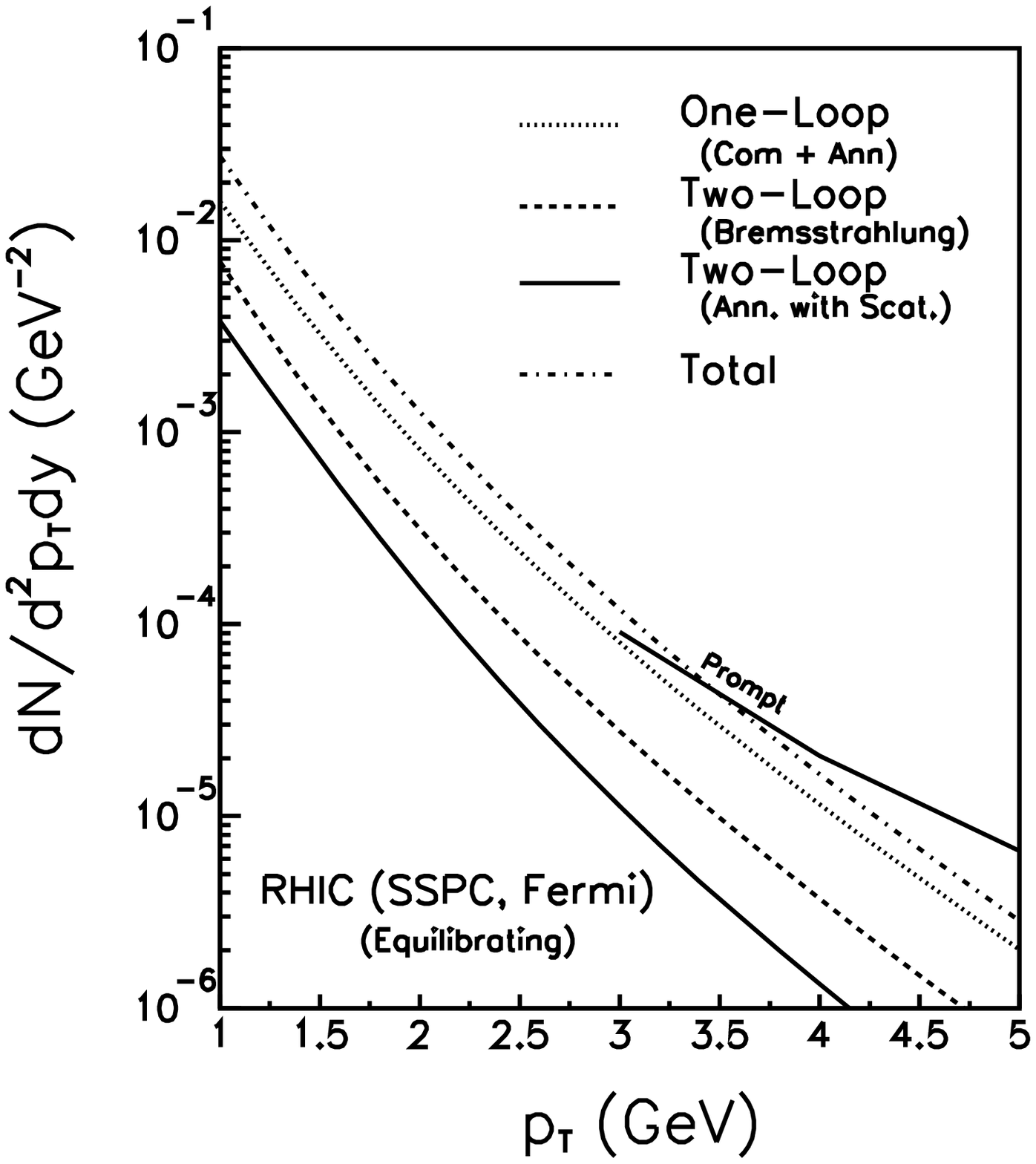,width=14cm,height=11cm}}
\caption{Photon spectra from various processes at RHIC energies 
with SSPC initial conditions and 
the Fermi-like profile function. The upper panel represents the fully 
equilibrated scenario, whereas the lower panel corresponds to the chemically 
equilibrating scenario.}
\end{figure}

\begin{figure}
\vspace*{-0.8cm}
\centerline{\psfig{figure=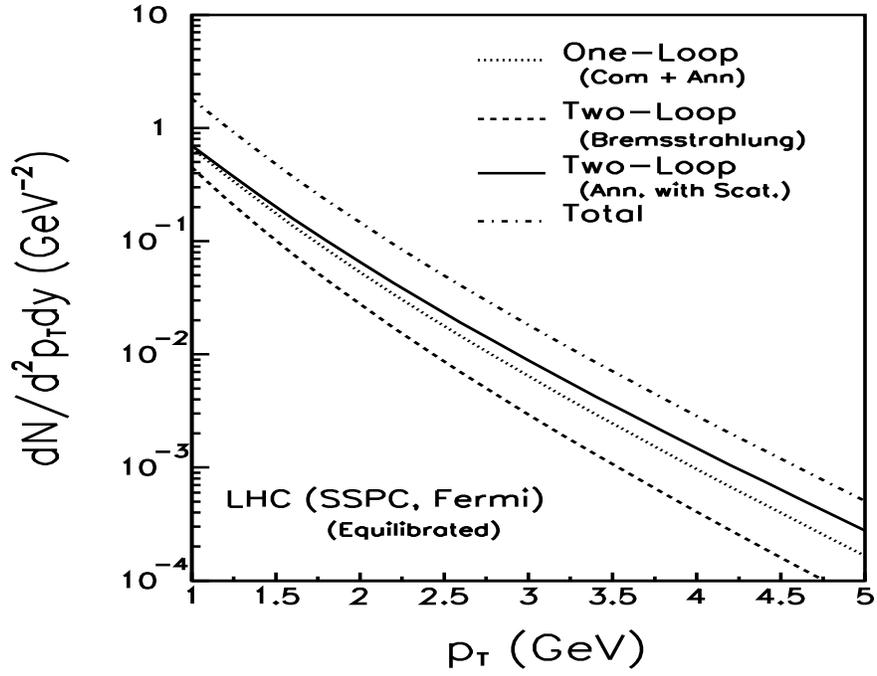,width=14cm,height=12cm}}
\vspace*{-1.9cm}
\centerline{\psfig{figure=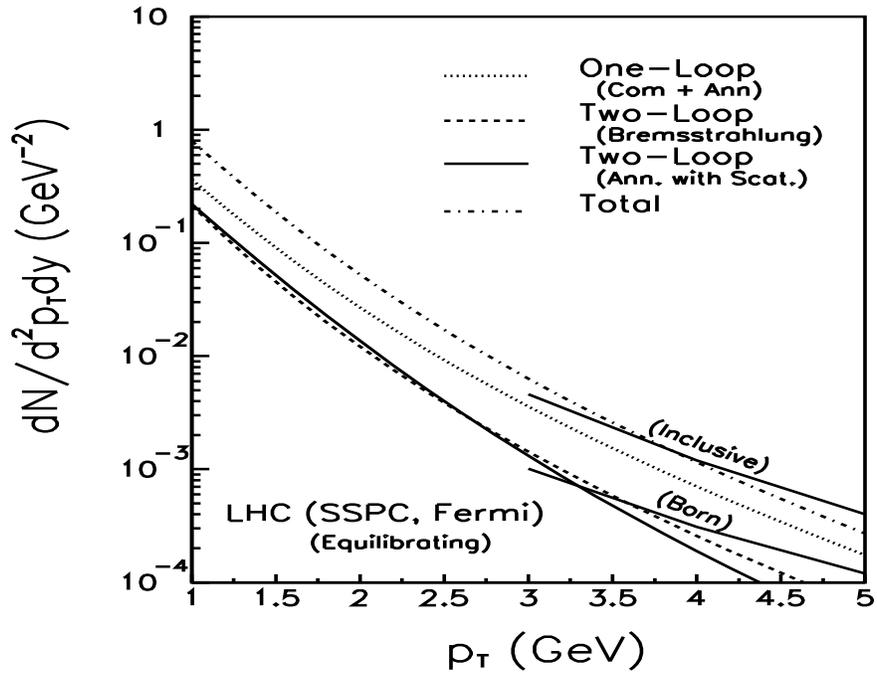,width=14cm,height=12cm}}
\caption{Same as Fig.1 for LHC energies.} 
\end{figure}

\begin{figure}
\vspace*{-0.8cm}
\centerline{\psfig{figure=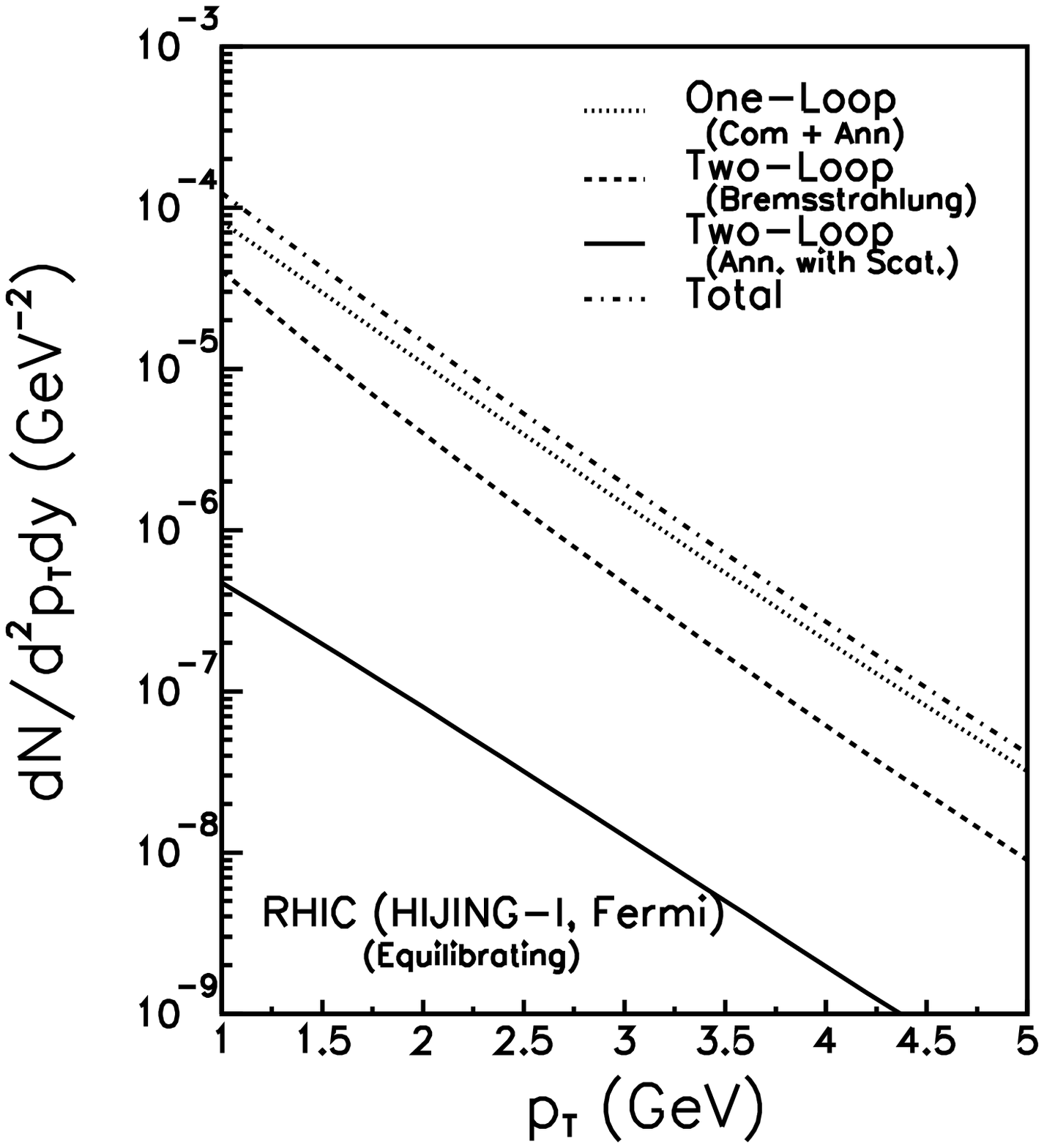,width=14cm,height=12cm}}
\vspace*{-1.9cm}
\centerline{\psfig{figure=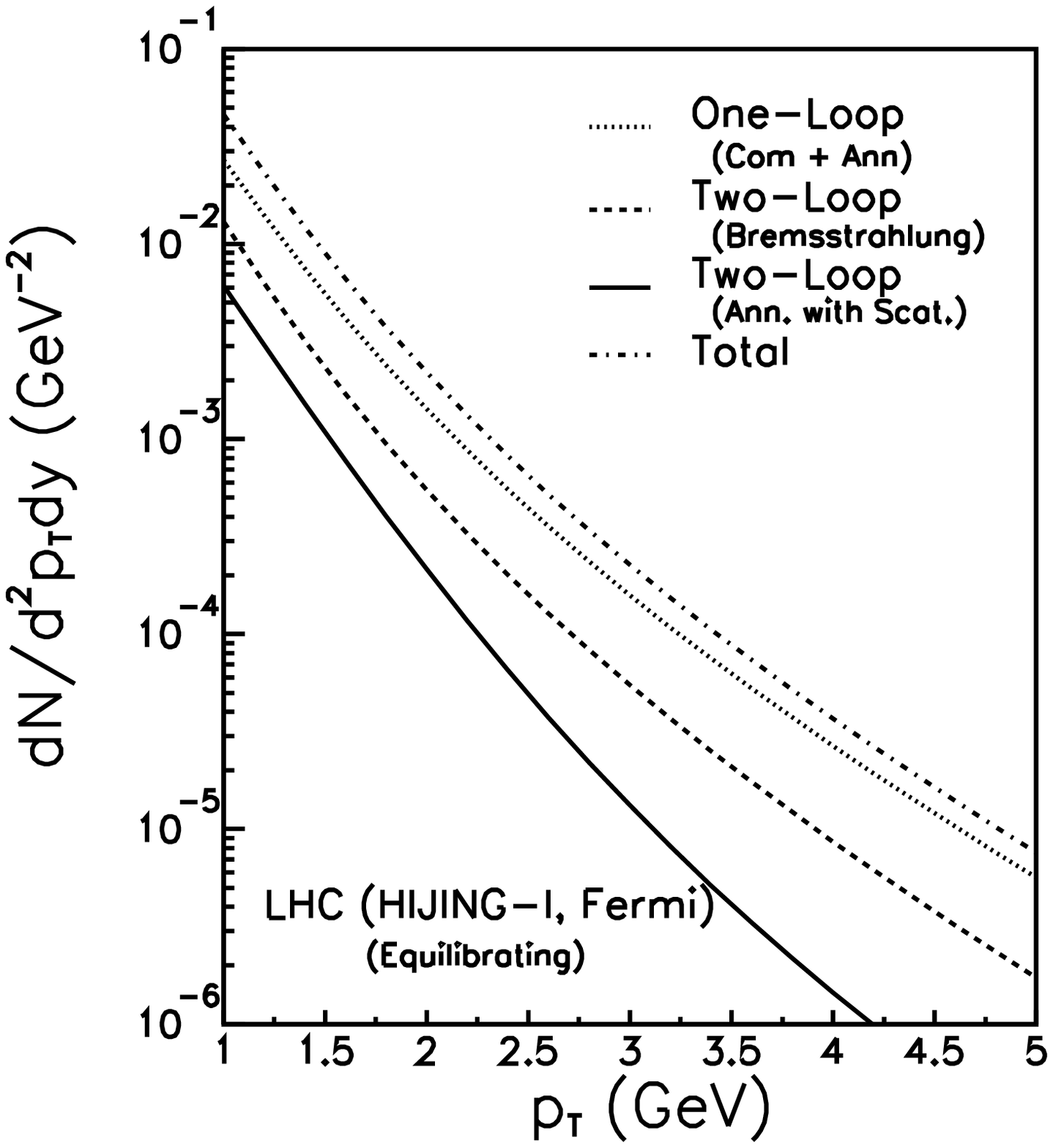,width=14cm,height=12cm}}
\caption{Photon spectra for a chemically equilibrating plasma
at RHIC (upper panel) and LHC (lower panel) energies
 with HIJING-I initial conditions and the Fermi-like profile function.} 
\end{figure}

\begin{figure}
\vspace*{-0.8cm}
\centerline{\psfig{figure=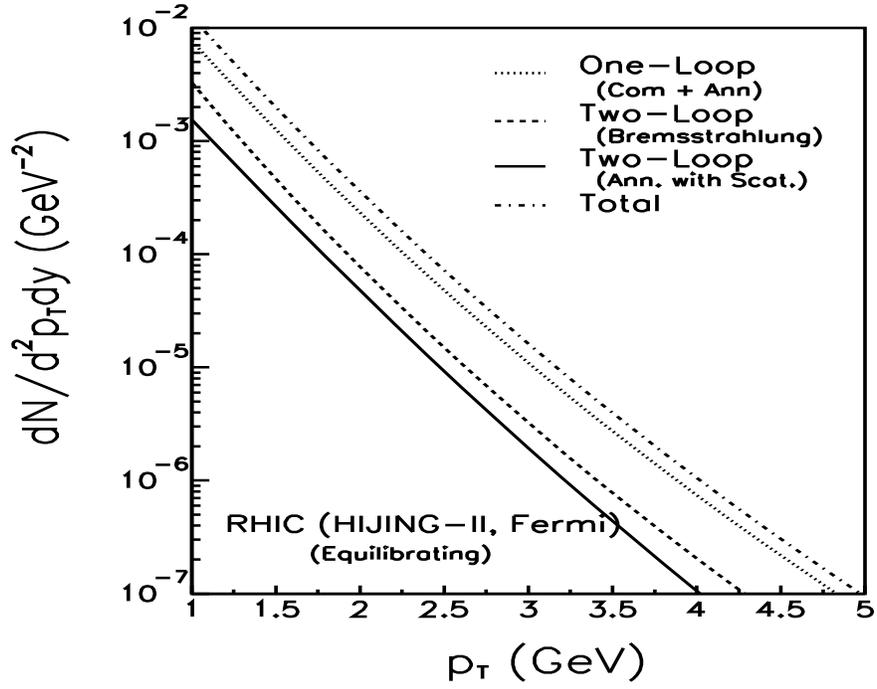,width=14cm,height=12cm}}
\vspace*{-1.9cm}
\centerline{\psfig{figure=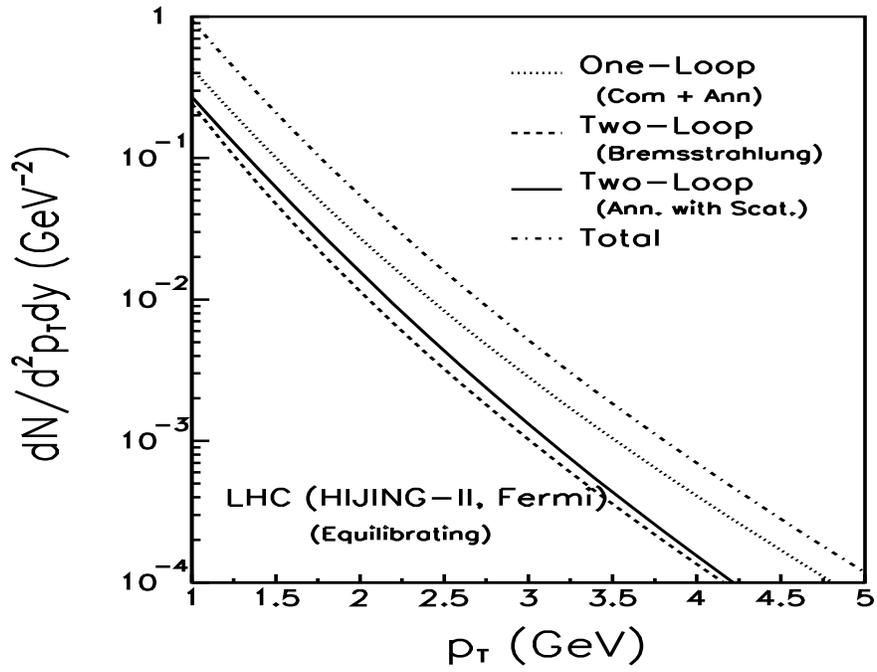,width=14cm,height=12cm}}
\caption{Same as Fig.3 with HIJING-II initial conditions.} 
\end{figure}

\end{document}